\documentclass[twocolumn]{aastex631}

\usepackage{multirow} 

\shorttitle{Simulation of collision-generated bar}
\shortauthors{Zhou et al.}
\graphicspath{{./}{figures/}}

\begin{document}

\title{N-body simulations of galaxy bars generated by satellite collisions: effects of the impact geometry}

\author[0009-0006-0194-6211]{Yufan Fane Zhou}
\email{yufanz@smail.nju.edu.cn}
\affiliation{School of Astronomy and Space Science, Nanjing University, Nanjing 210046, China}
\affiliation{Key Laboratory of Modern Astronomy and Astrophysics (Nanjing University), Ministry of Education, Nanjing 210046, China}

\author[0000-0003-0355-6437]{Zhiyuan Li}
\email{lizy@nju.edu.cn}
\affiliation{School of Astronomy and Space Science, Nanjing University, Nanjing 210046, China}
\affiliation{Key Laboratory of Modern Astronomy and Astrophysics (Nanjing University), Ministry of Education, Nanjing 210046, China}
\affiliation{Institute of Science and Technology for Deep Space Exploration, Suzhou Campus, Nanjing University, Suzhou 215163, China}

\author[0000-0002-1253-2763]{Hui Li}
\affiliation{Department of Astronomy, Tsinghua Univeristy, Beijing 100084, China}



\begin{abstract}
Bars, a common and important structure in disk galaxies, can be induced by galaxy interactions. Although there have been some studies on bar formation in flybys or collisions, the vast parameter space still leaves many scenarios that require further investigation. Here, we focus on the role of collisions caused by small galaxies (denoted as intruders), referred to as satellite collisions, in bar formation for MW/M31-like galaxies (denoted as target galaxies). Multiple sets of simulations with varying intruder initial velocities, inclination angles, collision positions, and intruder masses were run to study the dependence of this mechanism on these parameters. Our simulations show that bar formation favors moderate collision velocity, large inclination angle, off-center collision position, and large intruder mass. However, the bar's pattern speed and length are insensitive to these parameters, as the intruder's mass is relatively small compared to that of the target galaxy itself. Moreover, based on our tests, the intruder mass should be more than $\sim 3\times10^{9}$\,${\rm M}_{\odot}$ in order for this bar formation mechanism to operate effectively in MW/M31-like galaxies. Our work suggests the possibility that satellite collisions may have contributed, to some extent, to the bar formation in the Milky Way or M31.

\end{abstract}


\section{Introduction}
A significant number of disk galaxies are barred, although the bar fraction may vary depending on the samples and the definitions employed \citep{knapen2000,eskridge2000,menendez2007,sheth2008,cheung2013,buta2015}, and these bars have mainly been observed in optical and infrared wavelengths, not only in the local universe but also at high redshifts \citep[e.g.][]{jogee2004,guo2023,leconte2024}. Many studies suggest that bars play a crucial role in the evolution of disk galaxies, such as redistributing mass and angular momentum \citep{sakamoto1999,athanassoula2002}, triggering star formation \citep{ho1997,jogee2005}, (indirectly) fueling active galactic nuclei \citep{hopkins2010,galloway2015}, and building bulges or pseudobulges \citep{kormendy2004}.

The formation mechanisms of bars can be classified into two main categories: intrinsic and extrinsic \citep{noguchi1996}. On one hand, bars can form spontaneously in isolated galaxies due to dynamical instabilities inherent in self-gravitating axisymmetric disks \citep{efstathiou1982,athanassoula2013}. On the other hand, interactions with perturbers can induce the formation of bars in galaxies \citep{miwa1998}. For example, bars can form in satellite galaxies due to the tidal effects of their host galaxies \citep[e.g.][]{lokas2014}, in host galaxies due to the tidal effects of their satellites \citep[e.g.][]{mayer2004}, in galaxies orbiting a galaxy cluster due to the tidal effects of the cluster \citep[e.g.][]{mastropietro2005}, in pairs of galaxies with comparable masses due to their mutual flyby \citep[e.g.][]{lang2014}, as well as in galaxies that have experienced collisions or mergers due to these violent interactions \citep[e.g.][]{cavanagh2020}.

As a powerful tool, numerical simulation is widely utilized to investigate the formation of bars. However, despite the extensive simulations conducted, there are still gaps that remain to be filled. First, in tidal or merger scenarios, considering the vast parameter space of the interaction geometry, most studies have focused only on some representative orbital configurations. For instance, \citet{lang2014} and \citet{lokas2018} considered only planar prograde and retrograde interactions, neglecting intermediate inclination angles, to study the formation of tidally-induced bars in galaxy flybys. Second, as noted by \citet{cavanagh2020}, previous work mostly studied isolated bars and tidally-induced bars, with relatively few simulations focusing on bars induced by collisions or mergers. Third, simulations of bar formation in interactions between galaxies with extreme mass ratios are lacking. Although previous work \citep[e.g.][]{lang2014,lokas2018,cavanagh2020} have covered a broad range of mass ratios, from 1:1 to 10:1, the majority of satellite galaxies surrounding a host galaxy have masses less than 1/10 of the mass of the host. Taking the Milky Way (MW) as an example, there is only one satellite (the Large Magellanic Cloud) with a mass greater than 1/10 of its mass \citep{watkins2024}.

Therefore, here we aim to investigate the role of collisions caused by small galaxies (with a mass less than 1/10 of the target galaxy), referred to as satellite collisions, in the formation of bars. We are going to conduct multiple sets of numerical simulations involving a MW/M31-like galaxy and a small intruder, identifying whether a bar forms in the galactic disk after the collision. Interaction parameters, including impact angle, impact velocity, and impact position, will be varied to study how the geometry influences the formation of bars. The mass of the intruder will also be tested with various values to determine the minimum mass required for this mechanism to work. Additionally, gas may significantly influence bar formation \citep[e.g.][]{athanassoula2013_2}, so we will test several simulations that include gas as well.

The rest of the paper is structured as follows. Section~\ref{sec:methods} describes our galaxy models, simulation configurations and analysis techniques. The main results will be presented in Section~\ref{sec:results}, while additional results and discussions will be provided in Section~\ref{sec:discussions}. Section~\ref{sec:summary} offers a brief summary of our findings.

\section{Methods}
\label{sec:methods}
To simulate galaxy collisions, we first need to model both the target galaxy and the small intruder, generating isolated initial condition files for them. Then, for each set of simulations, we determine the interaction parameters (or orbital configurations) and combine the two galaxies into a single file, which is then run using the N-body simulation software \textsc{gadget-4} \citep{springel2005,springel2021}. To analyze the bar strength in our simulations, we will employ some mathematical methods.

\subsection{Galaxy models}
\label{subsec:models}
The target galaxy is modeled as a MW/M31-like galaxy, consisting of three N-body components \citep{Springel2005a}: an NFW \citep{navarro1996} dark matter halo, an exponential stellar disk and a Hernquist \citep{hernquist1990} stellar bulge. The structural parameters are similar to the M31 model in \citet{van2012}: the virial mass $M_{\rm vir}=1.5\times10^{12}$\,${\rm M}_{\odot}$, the halo concentration $c=9.56$, the disk mass $M_{\rm d}=1.2\times10^{11}$\,${\rm M}_{\odot}$, the disk scale length $r_{\rm s}=5$\,kpc, the disk scale height $h=1$\,kpc, the bulge mass $M_{\rm b}=1.9\times10^{10}$\,${\rm M}_{\odot}$, and the bulge scale length $a=1$\,kpc. The number of particles for halo, disk and bulge are 6,805,000, 12,000,000 and 1,900,000, as the particle masses for dark matter and stars are $2\times10^{5}$\,${\rm M}_{\odot}$ and $1\times10^{4}$\,${\rm M}_{\odot}$, respectively.

It is important to note that in the above model, the Toomre parameter is set to $Q=2$, which is higher than the normal value, preventing the galaxy from spontaneously forming a bar during its isolated evolution \citep[e.g.][]{lokas2018}. In a test simulation, the isolated galaxy does not form a bar within 8\,Gyr. Thus, we can ensure that, in collision simulations, the bar formed in the galaxy (if present) must be caused by the intruder.

As for the small intruder, considering its structure does not have a significant impact on the simulation results, we can assume it to be an NFW dark matter halo and a Hernquist stellar bulge. Its virial mass is set to $3\times10^{10}$\,${\rm M}_{\odot}$ and $6\times10^{10}$\,${\rm M}_{\odot}$ in different simulations, corresponding to 1/50 and 1/25 of the target galaxy's mass, respectively. Additionally, we briefly test some smaller masses (see Section~\ref{subsec:minimum}).

\subsection{Interaction parameters and simulation configurations}
\label{subsec:configurations}
\begin{figure}
    \centering
    \includegraphics[width=\columnwidth]{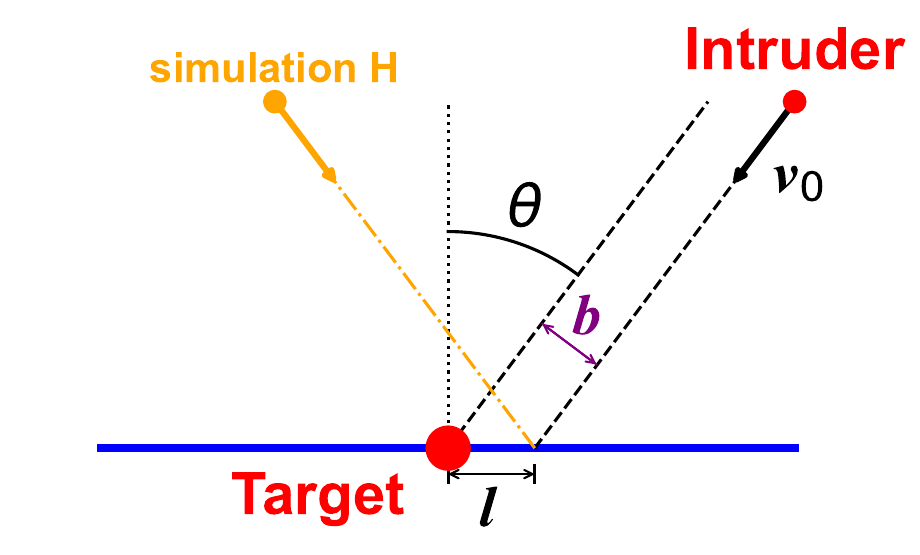}
    \caption{Schematic representation of the target galaxy and the intruder, showing $V_{\rm 0}$, $\theta$ and $l$ (the impact parameter $b$ can be derived from $\theta$ and $l$). The blue horizontal line represents the galaxy disk. Another geometric configuration of the off-center collision in Simulation~H is also presented, with details provided in Section~\ref{subsec:another}.}
    \label{fig:1}
\end{figure}

To study the effect of impact geometry on bar formation caused by satellite collisions, we focus on the following parameters (for a schematic illustration see Fig.~\ref{fig:1}, which is similar to Fig.~1 in \citet{fiacconi2012}). (\romannumeral1) The initial velocity $V_{0}$, which refers to the initial velocity of the intruder, with the target galaxy initially at rest. (\romannumeral2) The inclination angle $\theta$, which refers to the angle between the normal of the target galaxy and the initial velocity direction of the intruder. (\romannumeral3) The collision position $l$, which refers to the distance from the intersection of the intruder's initial velocity direction with the target galaxy's disk (`velocity-disk intersection') to the center of the disk. This intersection point may not be the actual collision point, but the difference between the two is small. Although the impact parameter $b$ is more commonly used, here $l$ offers a more intuitive indication of the impact location in the galaxy disk in the context of satellite collisions. (\romannumeral4) The intruder mass $m$. It is not a geometric parameter, but it is worth studying.

We run multiple sets of simulations with different values of $V_{\rm 0}$, $\theta$, $l$ and $m$. The initial parameters of our simulation suite are listed in Table~\ref{tab:1}. Initially, we place the center of mass of the target galaxy at the origin, with the disk lying in the $xy$ plane. The initial distance between the intruder and the `velocity-disk intersection' is 50\,kpc, and the velocity vector lies in the $xz$ plane. $V_{\rm 0}$ has three values: 250\,km\,s$^{-1}$, corresponding to a bound interaction; 500\,km\,s$^{-1}$ (the fiducial setting), corresponding to some fast interactions in a massive galaxy group \citep{carlberg2001}; and 1000\,km\,s$^{-1}$, which is twice the fiducial value. For $\theta$, we take values from 0\,deg to 90\,deg (edge-on collision), with a step of 10\,deg. $l$ has two values: 0\,kpc and 10\,kpc (off-center collision). The two values of $m$ are given in Section~\ref{subsec:models}.

\begin{table}
    \centering
    \begin{tabular}{ccccc}
    \hline
    Simulation & $V_{\rm 0}$ & $\theta$ & $l$ & $m$ \\
    label & (km\,s$^{-1}$) & (deg) & (kpc) & ($10^{10}$\,${\rm M}_{\odot}$) \\
    \hline
    A0 - A9 & 500 & 0 - 90 & 0 & 3 \\
    B0 - B9 & 250 & 0 - 90 & 0 & 3 \\
    C0 - C9 & 1000 & 0 - 90 & 0 & 3 \\
    D0 - D9 & 500 & 0 - 90 & 10 & 3 \\
    E0 - E9 & 500 & 0 - 90 & 0 & 6 \\
    \hline
    Fa & 500 & 90 & 0 & 0.5 \\
    Fb & 500 & 90 & 0 & 0.3 \\
    Fc & 500 & 90 & 0 & 0.1 \\
    Ga & 500 & 80 & 0 & 3 \\
    Gb & 500 & 90 & 0 & 3 \\
    H & 500 & 30 & 0 & 3 \\
    \hline
    \end{tabular}
    \caption{Simulation labels and their interaction parameters of 50 main simulations (upper part) and 6 extra simulations (lower part, see Section~\ref{sec:discussions} for details).}
    \label{tab:1}
\end{table}

Before the main simulations begin, we allow the target galaxy and the intruder to evolve in isolation for about 2\,Gyr to relax their initial condition files. Then, based on the interaction parameters, we combine the two files. We run each simulation for 5\,Gyr \citep[e.g.][]{lang2014} using the \textsc{gadget-4} code \citep{springel2021}, with a snapshot output every approximately 10\,Myr. The softening lengths for dark matter and stars are set to 30\,pc and 10\,pc, respectively.

\subsection{Fourier decomposition}
\label{subsec:fourier}
To quantify the bar strength, we perform a Fourier decomposition for the face-on stellar surface density of the target galaxy \citep{athanassoula2002}. The star particles we consider here are within a cylindrical region, centered on the galaxy's center, with a radius of 20\,kpc and a height of 10\,kpc (5\,kpc above and below the disk). The cylindrical region is divided into many concentric cylindrical shells, with a width d$R$ of 0.25\,kpc. For the cylindrical shell at radius $R$, $A_{\rm 2}$ is defined as the ratio of the second-order term to the zero-order term of the Fourier expansion:
\begin{equation}
    A_{\rm 2}(R)=\frac{|\Sigma_{\rm j}m_{\rm j}e^{2i\alpha_{\rm j}}|}{\Sigma_{\rm j}m_{\rm j}},
\end{equation}
where $m_{\rm j}$ is the mass of the $j$-th star particle and $\alpha_{\rm j}$ is its angular coordinate in the disk plane. The bar strength is characterized by $A_{\rm 2,max}$, which is the maximum value of $A_{\rm 2}(R)$ across all cylindrical shells. If a galaxy's $A_{\rm 2,max}\geq0.2$, it is considered to have a bar; if $A_{\rm 2,max}\geq0.3$, it is considered to have a strong bar \citep{rosas2020}. Here, the length of the bar ($R_{\rm A2,max}$) is defined as the value of $R$ where $A_{\rm 2}(R)$ reaches its maximum \citep[e.g.][]{rosas2020}. In addition, we can calculate the phase of the $m=2$ mode, $\Phi(R)$:
\begin{equation}
    \Phi(R)=\frac{1}{2}arctan[\frac{\Sigma_{\rm j}m_{\rm j}sin(2\alpha_{\rm j})}{\Sigma_{\rm j}m_{\rm j}cos(2\alpha_{\rm j})}],
\end{equation}
which should remain constant within the extent of the bar. To obtain the pattern speed of the bar, we just need to take the time derivative of $\Phi$.

\begin{figure*}
    \centering
    \includegraphics[width=\linewidth]{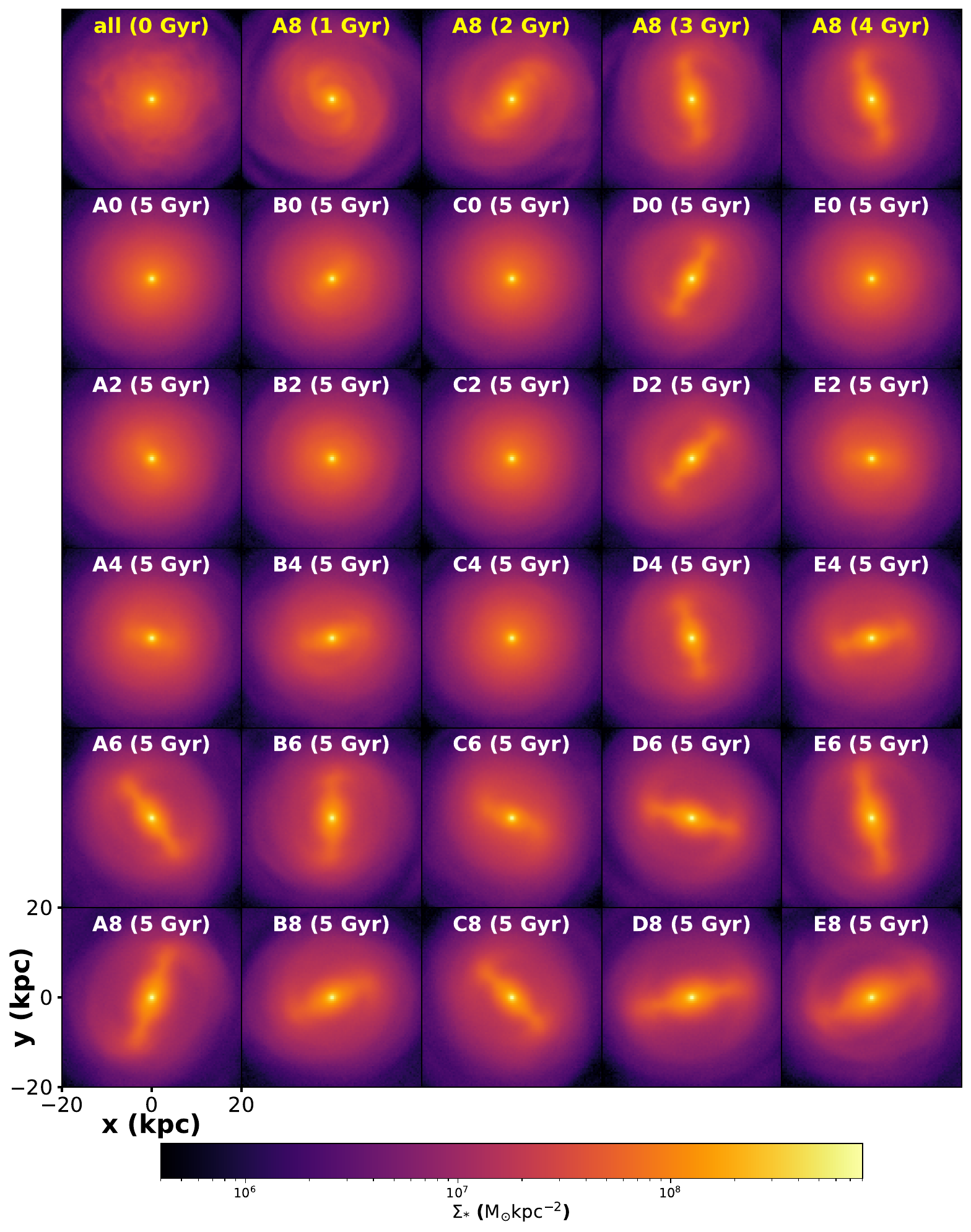}
    \caption{The stellar surface density of the target galaxy for certain snapshots. The collision happens at $\sim$80\,Myr. The first row shows the states of simulation~A8 at different times, while the remaining rows display the states of several main simulations at 5\,Gyr (limited by space, only showing simulations with $\theta$ of 0, 20, 40, 60 and 80\,deg).}
    \label{fig:2}
\end{figure*}

\begin{figure*}
    \centering
    \includegraphics[width=\linewidth]{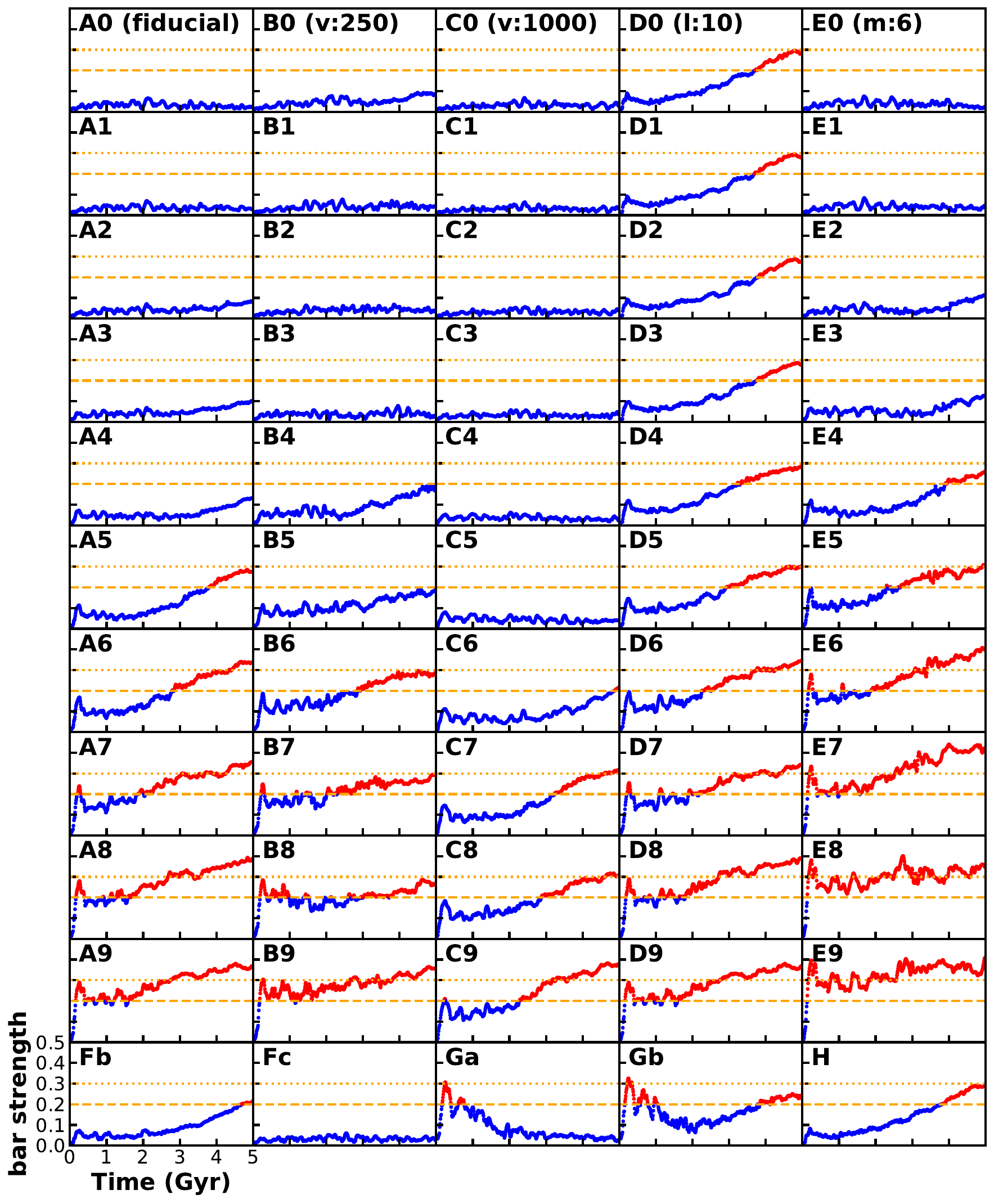}
    \caption{Evolution of the bar strength $A_{\rm 2}$ (defined in Section~\ref{subsec:fourier}) as a function of time for 50 main simulations and 5 extra simulations. Simulation labels are marked in the upper-left corners. The first row of Series B, C, D, and E shows their parameter differences relative to the fiducial series ($V_{\rm 0}$ changed to 250\,km\,s$^{-1}$, $V_{\rm 0}$ changed to 1000\,km\,s$^{-1}$, $l$ changed to 10\,kpc, and $m$ changed to $6\times10^{10}$\,${\rm M}_{\odot}$, respectively). The number in a label multiplied by 10 gives the inclination angle. The orange dashed (dotted) line corresponds to $A_{\rm 2}=0.2\,(0.3)$. For a data point, if its $A_{\rm 2}\geq 0.2$ (barred), it is plotted in red; otherwise, in blue.}
    \label{fig:3}
\end{figure*}
\section{Results}
\label{sec:results}
The collision caused by the small intruder will perturb the target galaxy, injecting asymmetries into its disk. If this perturbation is strong enough, a bar will form in the disk after a few gigayears. Fig.~\ref{fig:2} shows the stellar surface density for several simulations. For example, In simulation~A2, the disturbance brought by the collision is not significant, and no barred structure appears in the disk at 5\,Gyr (the end of the simulation). In contrast, the collision in simulation~D2 causes the disk to become asymmetric, ultimately leading to the formation of a bar.

Fig.~\ref{fig:3} shows the evolution of bar strength over time for all 50 main simulations. It is worth noting that in some simulations, in addition to the formation of a long-lived bar during the intermediate to late stages, the bar strength can also momentarily reach 0.2 or higher around 0.1\,Gyr (e.g. a red peak in A7 of Fig.~\ref{fig:3}). This transient enhancement is attributed to the collision occurring at $\sim 80$\,Myr, which significantly perturbs the galactic disk. When the collision geometry deviates substantially from a centrally symmetric configuration, such perturbations can immediately induce a transient bar-like structure, which, however, quickly dissipates. We are going to analyze the effects of different interaction parameters on bar formation (during the intermediate to late stages) based on Fig.~\ref{fig:3}.

First, we compare the simulation results from series A ($V_{\rm 0}=500$\,km\,s$^{-1}$), B ($V_{\rm 0}=250$\,km\,s$^{-1}$) and C ($V_{\rm 0}=1000$\,km\,s$^{-1}$) to study the effect of $V_{\rm 0}$. When $\theta\leq 40$\,deg, none of the simulations in the three series reach the bar threshold. For simulations with $\theta=50$\,deg, only A5 forms a bar. For larger inclination angles, such as 60\,deg, bars form earlier and with higher intensity in series A compared to B and C.  Such results suggest that a moderate relative velocity is most favorable for bar formation in satellite collisions. Slow collisions bring insufficient perturbation to the target galaxy, while fast collisions imply that the interaction time is too short. It it important to note that in series~B, the intruder is gravitationally bound to the target galaxy, so it passes through the disk multiple times. However, these multiple crossings do not make bar formation more easily than in series~A.

Second, we compare simulations within each series to analyze the effect of $\theta$. In simulations A, B, C and E, collisions with larger inclination angles lead to the formation of bars, while those with smaller angles do not. Additionally, the larger the inclination angle, the earlier the bar forms. Furthermore, although all simulations in series~D result in bar formation, the timing and the final strength follow the same trends as described above. Therefore, larger inclination angles favor the formation of bars in satellite collisions. 

Third, to investigate the role of $l$, series A ($l=0$\,kpc) and D ($l=10$\,kpc) are compared. When $\theta$ is small (less than 50\,deg), bar does not occur in A, but it does in D. For moderate $\theta$ (50\,deg and 60\,deg), the bar forms earlier in D than in A. While when $\theta$ is large (70\,deg and 80\,deg), the behaviors of A and D are similar. Clearly, off-center collisions are more likely to produce bars than central ones, as the former are geometrically highly asymmetric. However, when the inclination angle is very large, the effect of the collision position becomes less significant.

Last, let us study the effect of $m$ based on series A ($m=3\times10^{10}$\,${\rm M}_{\odot}$) and E ($m=6\times10^{10}$\,${\rm M}_{\odot}$). When $i=40$\,deg, simulation~A4 does not produce a bar, whereas simulation~E4 does. Additionally, for simulations with larger inclination angles, the bar forms earlier in series~E. A larger intruder mass facilitates the formation of a bar, which is intuitive in the context of satellite collisions. An unexpected situation is that in series~E, the intruder also reverses and performs a secondary crossing, which results in a mild oscillation in the bar strength after $\sim 3.5$\,Gyr. Ideally, the intruder's trajectory in E should be consistent with that in A, but due to differences in dynamical friction caused by varying masses, the behaviors diverge. However, the secondary crossing in E occurs at a late time and does not significantly affect the results.

Additionally, there are simulations (such as A4) that did not form a bar, yet their bar strength shows a steady increasing trend in Fig.~\ref{fig:3}. Perhaps with a longer runtime, these simulations would eventually form a bar \citep[e.g.][]{zheng2025b}. However, our goal is to explore which parameters are more likely to result in bar formation, and a duration of 5\,Gyr is already sufficient for this purpose.

$V_{\rm 0}$, $\theta$, $l$ and $m$ determine whether and when a bar forms after a satellite collision. However, the pattern speed of the bar seems to be insensitive to these parameters. Using the method in Section~\ref{subsec:fourier} to analyze the bars (if present) in all our simulations, we find that although their pattern speed may differ by the end, they are nearly identical at the time of formation. Bars undergo friction and decelerate over time, so the differences in their final speeds may merely reflect differences in their formation times. To further confirm the above point, we will study the pattern speed - bar strength ($\Omega_{\rm p}$-$A_{\rm 2}$) space \citep[e.g.][]{zheng2025} for the 27 simulations that ultimately formed a bar out of the 50 main simulations.

\begin{figure}
    \centering
    \includegraphics[width=\columnwidth]{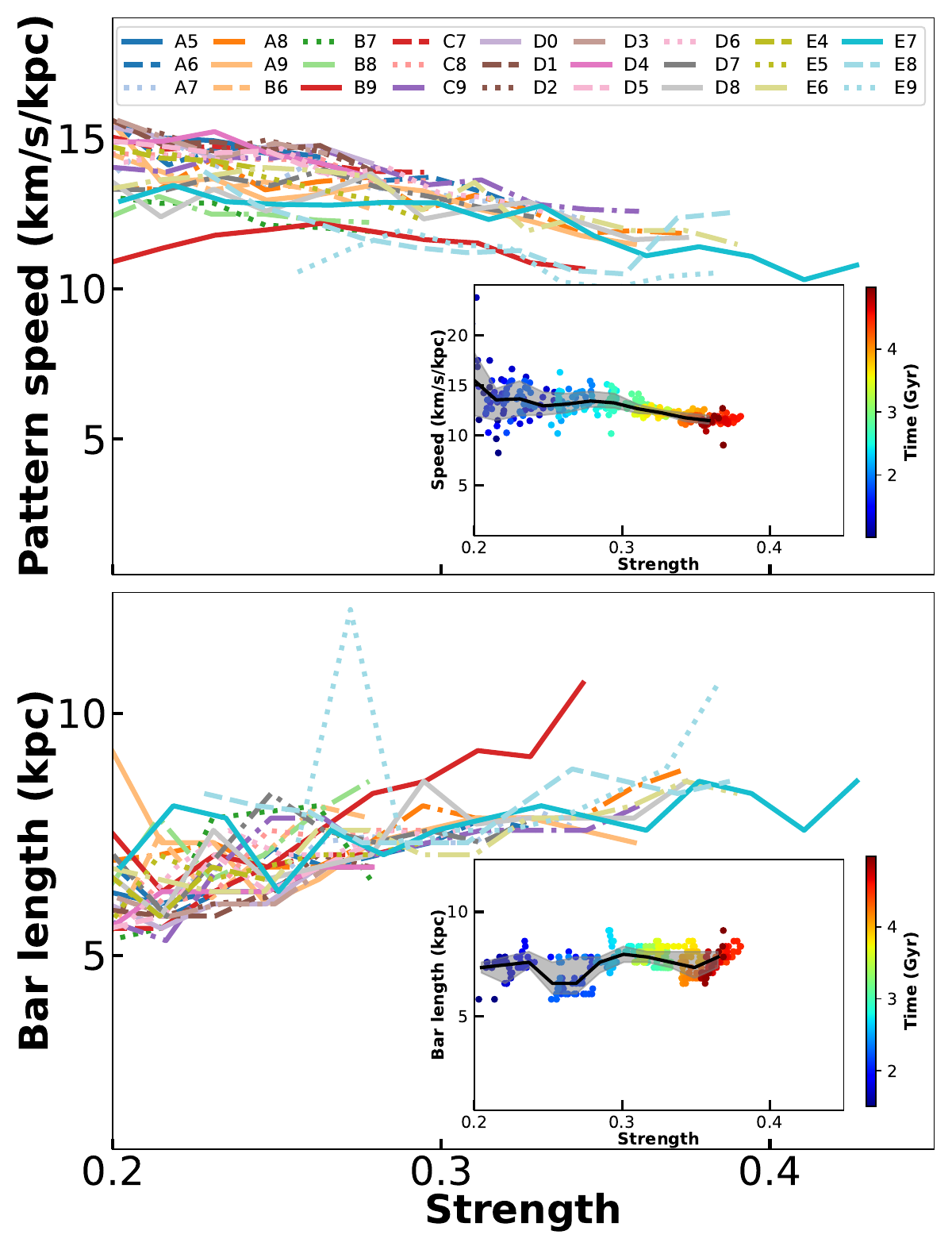}
    \caption{Upper panel: The $\Omega_{\rm p}$-$A_{\rm 2}$ plot for 27 main simulations in which a bar is formed. The inset, using Simulation~A9 as an example, illustrates how each curve in the main panel is constructed: each scatter point corresponds to a snapshot in the simulation and is color-coded by time. These points are then binned by bar strength $A_{\rm 2}$, and the $\Omega_{\rm p}$-$A_{\rm 2}$ curve is drawn by connecting the median value within each bin. Lower panel: The $R_{\rm A2,max}$-$A_{\rm 2}$ plot, constructed in a manner similar to the upper panel.}
    \label{fig:4}
\end{figure}

For a given simulation, each snapshot corresponds to a data point on the $\Omega_{\rm p}$-$A_{\rm 2}$ plane. We bin these data points based on their $A_{\rm 2}$ values, and then construct the $\Omega_{\rm p}$-$A_{\rm 2}$ curve by plotting the median $\Omega_{\rm p}$ value within each bin. The upper panel's inset of Fig.~\ref{fig:4} demonstrates the above procedure using simulation~A9 as an example, with the shaded region marking the 16th and 84th percentiles, corresponding to the 1$\sigma$ range. All curves of the 27 simulations are presented in the upper panel of Fig.~\ref{fig:4}, with most occupying the same region in the $\Omega_{\rm p}$-$A_{\rm 2}$ space. This suggests that the pattern speed of the bar triggered by satellite collision is indeed insensitive to interaction parameters, due to the intruder's mass being relatively small compared to that of the target galaxy. It is worth noting that several simulations (e.g. B7, B8, B9, E8, E9) exhibit curves that deviate from the majority. These cases correspond to bound encounters with large inclination angles, where repeated passages of the intruder severely disrupt the normal evolution of the bar, as shown in Fig.~\ref{fig:3}.

Whether the bar length $R_{\rm A2,max}$ is related to these parameters can also be investigated using a similar approach. The $R_{\rm A2,max}$-$A_{\rm 2}$ plot is shown in the lower panel of Fig.~\ref{fig:4}, where it can be seen that the bar length increases with increasing bar strength. Except for a few scattered cases from high-inclination bound simulations (e.g. B9 and E9), the overall consistency of the curves in $R_{\rm A2,max}$-$A_{\rm 2}$ space suggests that the bar length is also insensitive to the parameters in satellite collisions. Note that besides $R_{\rm A2,max}$ adopted in this study, there are some alternative proxies for bar length (e.g. $R_{\rm \Phi}$, the radii within which the phase $\Phi$ remains constant). These alternative definitions could be larger than $R_{\rm A2,max}$ \citep{rosas2020,anderson2024}, so adopting other proxies may shift all curves in the lower panel of Fig.~\ref{fig:4} upward as a whole. However, they would still occupy the same region, leaving our conclusions unaffected.

\section{Discussions}
\label{sec:discussions}
\subsection{The minimum mass of the intruder}
\label{subsec:minimum}
Since the satellite collision is a viable mechanism for bar formation, we hope to further investigate the minimum intruder mass required for this process to be effective. The other three parameters are set to the most favorable conditions for bar formation, i.e., $V_{\rm 0}=500$\,km\,s$^{-1}$, $\theta=90$\,deg and $l=10$\,kpc, based on the conclusions in Section~\ref{sec:results}. We gradually decrease the value of $m$ and conduct multiple simulation tests. It is found that when $m=5\times10^{9}$\,${\rm M}_{\odot}$, a relatively strong bar still appears in the simulation (denoted as Fa). In the case of $m=3\times10^{9}$\,${\rm M}_{\odot}$, a weak bar only forms at the end of the simulation (denoted as Fb, see Fig.~\ref{fig:3}). However, when $m$ is reduced to $1\times10^{9}$\,${\rm M}_{\odot}$, no bar forms, and the bar strength shows almost no increase over the 5\,Gyr simulation period (denoted as simulation~Fc, see Fig.~\ref{fig:3}).

Therefore, in order to form a bar through a satellite collision for a MW/M31-like galaxy, the mass of the small intruder needs to be at least $\sim3\times10^{9}$\,${\rm M}_{\odot}$. Of course, here we are merely discussing this mechanism and not suggesting that the bar of a specific galaxy \citep[e.g. M31,][]{feng2022,feng2024} is necessarily caused by an intruder with a mass that meets the conditions. However, considering that both the Milky Way and M31 host several satellite galaxies with masses greater than $3\times10^{9}$\,${\rm M}_{\odot}$, it is indeed plausible that satellite collisions have played a role in the bar formation of the Milky Way and M31.

\subsection{The role of gas}
\label{subsec:gas}
\citet{athanassoula2013_2} indicated that gas can help the disk stay near-axisymmetric for a longer period, preventing the formation of a bar. Moreover, in gas-rich galaxies, both the growth rate and final strength of the bar are lower than those in gas-poor galaxies. To investigate whether the above conclusion holds in the context of a satellite collision, we convert 30\% of the stellar particles in the target galaxy's disk into gas particles and rerun the 9 simulations of series~A, employing the effective equation of state for gas colder than $10^4$\,K and star formation \citep{springel2002,springel2003}, with radiative cooling modeled through collisional ionization equilibrium under a spatially constant but time-variable UV background \citep{springel2021}.

It is found that the presence of gas significantly suppresses bar formation induced by a satellite collision: even at a high inclination angle of $\theta=80$\,deg, no bar develops within 5\,Gyr, and in the extreme case of $\theta=90$\,deg, a bar only appears near the end of the simulation (denoted as simulations Ga and Gb, see Fig.~\ref{fig:3}). This presents a stark contrast to the simulations of Series~A and is consistent with \citet{athanassoula2013_2}.

\subsection{Another geometric configuration in the off-center case}
\label{subsec:another}
When $l=0$\,kpc, the effects of an intruder (with a fixed $\theta$) colliding from different directions should, in theory, be identical due to the symmetry of the system. However, for off-center cases, collisions from different directions are not geometrically equivalent (e.g. the two scenarios in Fig.~\ref{fig:1}). As a simple test, we alter the collision direction of the intruder in simulation~D3 in a new simulation (denoted as H) to explore the influences of such geometric configuration.

By comparing the panel H with D3 in Fig.~\ref{fig:3}, it can be seen that the collision direction has little effect on bar formation in satellite collisions. In other bar formation mechanisms, different geometric configurations may lead to distinctly different outcomes. For example, prograde equal-mass coplanar flybys are more likely to form bars than retrograde equal-mass coplanar flybys \citep[e.g.][]{lokas2018}. However, in the case of satellite collisions, the intruder's mass is too small, and thus, given fixed interaction parameters, the direction of the collision is not significant.

\section{Summary}
\label{sec:summary}
Collisions with small intruders can lead to bar formation in the target galaxy. To study the dependence of this mechanism on interaction parameters, including the intruder's initial velocity $V_{\rm 0}$, inclination angle $\theta$, collision position $l$, and intruder mass $m$,  we performed a series of N-body simulations using \textsc{gadget-4} \citep{springel2021}. Fifty main simulations include 3 values of $V_{\rm 0}$ (500\,km\,s$^{-1}$, 250\,km\,s$^{-1}$ and 1000\,km\,s$^{-1}$), 10 values of $\theta$ (0\,deg, 10\,deg, ..., 90\,deg), 2 values of $l$ (0\,kpc and 10\,kpc), and 2 values of $m$ ($3\times10^{10}$\,${\rm M}_{\odot}$ and $6\times10^{10}$\,${\rm M}_{\odot}$), with each simulation running for 5\,Gyr. The properties of bars are analyzed with Fourier decomposition.

Our results show that bar formation in satellite collisions favors: (\romannumeral1) moderate initial velocity, (\romannumeral2) large inclination angle, (\romannumeral3) off-center collision position, and (\romannumeral4) large intruder mass. However, these parameters do not affect the pattern speed and length of the resulting bar, as the intruder's mass is small.

Several additional simulations were conducted to further investigate the effects of intruder mass, gas, and geometric configuration. For a MW/M31-like galaxy, the intruder's mass must be at least $3\times10^{9}$\,${\rm M}_{\odot}$ (for the extreme case of $\theta=90$\,deg) for the satellite collision mechanism to work. We also found that gas in the target galaxy can inhibit bar formation, consistent with the conclusions of previous simulations. Moreover, the collision direction has little impact in off-center cases, again due to the small mass of the intruder.

We explores the interaction parameter space for satellite collisions, a bar formation mechanism that has not been extensively studied before. The results indicate that satellite collisions may have played a role in the bar formation of the Milky Way and M31. Our work provides a complement to previous theoretical and simulation studies, contributing to a deeper understanding of bar formation in galaxies.

\begin{acknowledgments}
We thank Juntai Shen and Shihong Liao for helpful discussions. This work is supported by the National Natural Science Foundation of China (grant No. 12225302) and the National Key Research and Development Program of China (grant No. 2022YFF0503402), and the China Manned Space Program with grant No. CMS-CSST-2025-A10. H.L. is supported by the National Key R\&D Program of China No. 2023YFB3002502, the National Natural Science Foundation of China under No. 12373006, and the China Manned Space Program with grant No. CMS-CSST-2025-A10.
\end{acknowledgments}

%

\vspace{5mm}


\software{GADGET-4 \citep{springel2005,springel2021}}







\bibliography{bar}{}
\bibliographystyle{aasjournal}



\end{document}